\documentclass[letterpaper,12pt]{article}
\usepackage{amsfonts}
\usepackage{amsmath,amsthm}
\usepackage{amssymb,amscd}
\usepackage{enumerate}

\makeatletter
\newcommand{\rmnum}[1]{\romannumeral #1}
\newcommand{\Rmnum}[1]{\expandafter\@slowromancap\romannumeral #1@}
\makeatother

{}
\newtheorem{theorem}{Theorem}[section]

\newtheorem{lemma}[theorem]{Lemma}
\newtheorem{proposition}[theorem]{Proposition}

\theoremstyle{definition}

\theoremstyle{remark}

\swapnumbers \makeatletter
\def\ps@mystyles{    \def\@evenhead{\hfil{\sl\leftmark}\hfil}    \def\@oddhead{\hfil{\sl\rightmark}\hfil}    }
\makeatother

\begin{document}
\title{Transition Probabilities of the Bethe Ansatz Solvable Interacting Particle Systems}

\author{\textbf{Eunghyun Lee}\\ \small{\textit{Department of Mathematics}}
                             \\ \small{\textit{University of California}}
                             \\ \small{\textit{Davis, CA 95616, USA}}
 \date{}                            \\ \small{\textit{Email:ehnlee@math.ucdavis.edu}}}
\maketitle

\begin{abstract} \noindent This paper presents the exact expressions of the transition probabilities of some \textit{non-determinantal} Bethe ansatz solvable interacting particle systems: the two-sided PushASEP, the asymmetric avalanche process and the asymmetric zero range process.  The time-integrated currents of  the asymmetric avalanche process and the asymmetric zero range process are immediate from the results of the asymmetric simple exclusion process.
\end{abstract}

\section{Introduction}
The Bethe ansatz technique is one of the methods widely used to find exact solutions of interacting particle systems.  In physics the Bethe ansatz technique was originally  used to obtain the eigenvalues and the eigenvectors of the Hamiltonian of the one-dimensional quantum spin-1/2 chain \cite{Sutherland}. However, from the fact that Markov matrices of some interacting particle systems are a similarity transformation of the Hamiltonians of quantum spin chains one can study dynamical properties of the interacting particle systems by using the Bethe ansatz \cite{Gwa}. \\
\indent Among  the Bethe ansatz solvable models, the asymmetric simple exclusion process (ASEP)  is a paradigmatic model in non-equilibrium statistical mechanics.  A sub-model of the ASEP, the totally  asymmetric simple exclusion process (TASEP) has been connected to random matrix theory due to the seminal works \cite{Baik,Johannson} (for review, see \cite{Ferrari,Krug}) and some results on the TASEP have recently been extended to the ASEP [24--28]. In this direction, the mapping between the TASEP and the corner growth model based on the RSK algorithm and the Bethe ansatz technique have been mainly used. Sch\"{u}tz \cite{Schutz} found the transition probability of the TASEP with $N$ particles and expressed it as an $N \times N$ determinant whose entries are hypergeometric functions. This exact solution of the Kolmogorov forward equation is a starting point of time-integrated currents of the system and related works. Indeed, the current distribution of the TASEP and its large time asymptotics were well studied by using the determinant representation and some properties of entries of the determinant \cite{Borodin,Imamura,Nagao,Rakos,Sasamoto}.  This determinantal structure of the TASEP is inherited from the $S$-matrix that describes a two-particle interaction in the TASEP.  The $S$-matrix in the ASEP is
 \begin{equation}
   S_{\beta\alpha} = -\frac{p+q\xi_{\alpha}\xi_{\beta} - \xi_{\beta}}{p+q\xi_{\alpha}\xi_{\beta} - \xi_{\alpha}},~~~(\xi_{\alpha}, \xi_{\beta} \in \mathbb{C}). \label{scattering}
 \end{equation}
  In case of the TASEP, that is $p=1$ and $q=0$,  (\ref{scattering}) is \textit{separable} in the sense that the numerator  has only $\xi_{\beta}$ variable and the denominator has only $\xi_{\alpha}$ variable. This separability of variables makes it possible to express the transition probability as a determinant. However, as one sees, the ASEP does not have this property and thus the determinantal structure is no longer in the ASEP.\\
    \indent The determinantal structure appears in the drop-push model \cite{Sasamoto2}, too. In the drop-push model a particle can jump to the right neighboring site with rate 1 even if it is already occupied, pushing all the right neighboring particles by one. That is, representing a particle by $A$ and an empty site by $0$, the process
  \begin{equation}
   \underbrace{A\cdots A}_n0 \rightarrow 0\underbrace{A \cdots A}_n \label{pushingR}
  \end{equation}
  occurs with rate 1 for all $n \in \mathbb{N}$ when the exponential clock of the leftmost particle in (\ref{pushingR}) rings.  Even if we allow particles to jump in both directions,  the resulting process is still determinantal if the pushing effect is one-sided. Indeed, Borodin and Ferrari \cite{Borodin2} have shown that limiting processes are the $\textrm{Airy}_1$ process and $\textrm{Airy}_2$ process for  flat initial condition and  step initial condition, respectively. For the ASEP, which is not determinantal, in a breakthrough [24--26] Tracy and Widom obtained the distribution of time-integrated current and its large-time asymptotics for step initial condition. Tracy and Widom found first the transition probability of the ASEP with $N$ particles  and then obtained an integral formula for the distribution of time-integrated current by summing the transition probability over all possible configurations. A novel point of Tracy and Widom's transition probability is that it is given by an integral representation over arbitrarily  small contours or large contours keeping the forms of $S$-matrices in the integrand. Thus, one can find the probability distribution of the $m$th particle's position by using various combinatorial identities related to the $S$-matrix of the ASEP.
\\
  \indent In this paper we find the exact expression of transition probabilities of other Bethe ansatz solvable interacting particle systems which are not determinantal. The transition probabilities are a starting point of working on the currents of systems as we saw in Tracy and Widom's works. The models we will consider are the two-sided PushASEP, the asymmetric avalanche process (ASAP) and the asymmetric zero range process (AZRP) with constant rates. We shortly introduce the models. If the process (\ref{pushingR}) occurs with rate $r_n$ dependent on $n$, the process is no longer determinantal, and  in this case, $r_n$ is fixed to
  \begin{equation}
  r_{n} = \frac{1}{1 + {\lambda}/{\mu} + ({\lambda}/{\mu})^2 + \cdots +  ({\lambda}/{\mu})^{n-1}},\hspace{0.5cm}(\lambda + \mu=1) \label{RightRate}
  \end{equation}
   to permit the Bethe ansatz solvability \cite{Ali}. This model which is called the \textit{generalized totally asymmetric exclusion process} or the \textit{one-parameter family of asymmetric exclusion process} was first suggested by Alimohammadi, Karimipour and Khorrami \cite{Ali}. The \textit{generalized totally asymmetric exclusion process} interpolates between the TASEP ($\lambda \rightarrow 1$) and the drop-push model ($\lambda \rightarrow 0$). Later this model was extended to the asymmetric case called the \textit{two-parametric family of asymmetric exclusion process} \cite{Ali2}. In the asymmetric case the process is
  \begin{equation}
   \underbrace{A\cdots A}_n0  \rightleftharpoons  0\underbrace{A \cdots A}_n \label{BothPushing}
  \end{equation}
  with rates $pr_n$ to the right and $ql_n$ to the left where $r_n$ is (\ref{RightRate}) and
  \begin{equation}
  l_{n} = \frac{1}{1 + {\mu}/{\lambda} + ({\mu}/{\lambda})^2 + \cdots +  ({\mu}/{\lambda})^{n-1}},\hspace{0.5cm}(\lambda + \mu=1). \label{LeftRate1}
  \end{equation}
  When $p,q \neq 0$, this asymmetric model interpolates between the model with the pushing dynamics on both sides ($\lambda,\mu \neq 0$) and the model \cite{Borodin2} with the TASEP dynamics on one side and the pushing dynamics on the other side ($\lambda$ (or $\mu$) = 1).
   In this paper we will call this \textit{two-parametric family of asymmetric exclusion process} simply the \textit{two-sided PushASEP} and assume that $\lambda$ and $\mu$ are nonzero. \\
   \indent  In the ASAP \cite{Povol1,Povol2} particles jump to the right or to the left with rate $p$ and $q$, respectively, after an exponential time. When particles try to jump to the site already occupied,  an avalanche starts immediately. The rule for the avalanche is as follows\footnote{It takes an infinitesimal time for the avalanche to end.}: If $n~(\geq 2)$ particles are at site $x$ during an infinitesimal time, then $n$ particles jump to $x+1$ all together with probability $\mu_n(\mu_n\neq 1)$ or with probability $\lambda_n$, $n-1$ particles jump to $x+1$ all together and one particle stays at $x$  where $\mu_n + \lambda_n =1$. For example, for a two-particle system with $p=1,q=0$, if particles are initially at $x$ and $x+1$ and the exponential clock at $x$ rings at time $t$, then the resulting configuration is one of $(x+n,x+n+1),~ n \in \mathbb{N}$ and the probability of finding two particles at $x+n$ and $x+n+1$ is $\lambda\mu^{n-1}$ where $\mu=\mu_2$ and $\lambda= \lambda_2$. Regarding the Bethe ansatz solvability, it is known \cite{Povol1,Povol2} that $\mu_n$ must be given by
\begin{equation}
 \mu_n = \mu\frac{1-(-\mu)^{n-1}}{1+\mu},~~n \geq 2. \label{Constraint}
\end{equation}
As a special case, if $\mu=-p/q$, the process becomes the ordinary ASEP \cite{Povol2}, and it is easily seen that the ASAP with $\mu=0$ is equivalent to the two-sided PushASEP with $\mu=1$ and so the process is determinantal. In this paper we consider only the case $0< \mu <1$. \\
   \indent The asymmetric zero range process (AZRP) is also not determinantal ($p,q\neq 0$).  The dynamics of general AZRP is governed by the following laws; if a site $x$ is occupied by $n$ particles, one of the particles at $x$ leaves the site $x$ at the rate $g(n)$. In other words, the jumping rates $g(n)$ depend on the number of particles at the departure site.  The particle that leaves $x$ chooses a target site $y$ with probability $p(x,y)$. In this paper, we assume that  $p(x,x+1) =p$ and $p(x,x-1) =q$ with $p+q=1$ and $g(n)=1$  for all $n$.
\\
\indent The paper is organized as follows. In section 2, we discuss the forward equations and the $S$-matrices of the models as a background. In section 3,  we provide the main result, the transition probabilities (\ref{transition}) of the models, and  using the obtained transition probabilities, we revisit the map between configurations of the AZRP  and  configurations of  the ASEP \cite{Evans,Harris,Kipnis2,Timo}. Also, we shortly discuss the current distributions of the ASAP and the AZRP.
\section{Preliminaries}
In this section we review the standard method by the idea of Bethe  to discuss the $S$-matrices of the models that characterize each model.
The Bethe ansatz solvable models in this paper with a finite number of particles are continuous-time Markov processes with countable state spaces. A configuration of the models is specified by each particle's position and we denote the configuration by $X= (x_1,\cdots,x_N) \in \mathbb{Z}^N$. Let $P_Y(X;t)=P_Y(x_1,\cdots,x_N;t)$ be the transition probability from the initial configuration $Y$ to a configuration $X$ at time $t$ and we call the domain of $P_Y(X;t)$ as a function of $X$ the \textit{physical region} of the model. The physical region of the two-sided PushASEP and the ASAP is
\begin{equation*}
 \{(x_1,\cdots,x_N) \in \mathbb{Z}^N : x_1 < \cdots < x_N\}
\end{equation*}
and the physical region of the AZRP is
\begin{equation*}
 \{(x_1,\cdots,x_N) \in \mathbb{Z}^N : x_1 \leq  \cdots \leq x_N\}.
 \end{equation*}
  Depending on $X$, the $P_Y(X;t)$ satisfies different forward equations. For example, in the TASEP with two particles, $P_Y(x_1,x_2;t)$ with $x_1 < x_2-1$ and $P_Y(x_1,x_2;t)$ with $x_1 = x_2-1$ satisfy
 \begin{equation}
 \frac{d}{dt}~P_Y(x_1,x_2;t) = P_Y(x_1 -1,x_2;t) + P_Y(x_1,x_2-1;t) - 2P_Y(x_1,x_2;t) \label{master1}
 \end{equation}
 and
 \begin{equation}
 \frac{d}{dt}~P_Y(x_1,x_2;t) = P_Y(x_1 -1,x_2;t)  - P_Y(x_1,x_2;t),\label{master2}
 \end{equation}
  respectively.  However, introducing a function $u(x_1,x_2;t)$ on $\mathbb{Z}^2 \times [0,\infty)$ that satisfies (\ref{master1}) for noninteracting particles and a \textit{boundary condition} at the boundary of the physical region,  two differential equations (\ref{master1}) and (\ref{master2}) are combined as a single differential equation with the boundary condition. That is, for $u(x_1,x_2;t)$ on $\mathbb{Z}^2 \times [0,\infty)$,
  \begin{equation}
 \frac{d}{dt}~u(x_1,x_2;t) = u(x_1 -1,x_2;t) + u(x_1,x_2-1;t) - 2u(x_1,x_2;t) \label{master3}
 \end{equation}
 with
 \begin{equation}
  u(x,x;t) = u(x,x+1;t) \label{boundary0}
 \end{equation}
 is in the form of (\ref{master1}) when   $x_1<x_2-1$ and in the form of (\ref{master2}) when $x=x_1 = x_2-1.$  Hence, if $u(x_1,x_2;t)$ satisfies (\ref{master3}), (\ref{boundary0}) and the initial condition
 \begin{equation*}
 u(x_1,x_2;0) = \delta_{y_1}(x_1)\delta_{y_2}(x_2) \hspace{0.2cm}\textrm{when}\hspace{0.2cm}x_1<x_2,
 \end{equation*}
 then
  $u(x_1,x_2;t) = P_Y(X;t)$ with $X=(x_1,x_2)$ and $Y=(y_1,y_2)$ in the physical region. As a Bethe ansatz solvable model, the TASEP with $N$ particles ($N \geq3$) does not need new constraints except the one in the form of (\ref{boundary0}). Hence the solution of
    \begin{equation}
 \frac{d}{dt}~u(X;t) = \sum_{i=1}^N\Big(u(x_1,\cdots,x_{i-1},x_i-1,x_{i+1},\cdots,x_N;t) - Nu(x_1,\cdots,x_N;t)\Big) \label{totallyMaster}
\end{equation}
with the boundary condition
  \begin{equation}
u(x_1,\cdots,x_i,x_i+1,\cdots,x_N;t) = u(x_1,\cdots,x_i,x_i,\cdots,x_N;t) \label{boundaryTASEP}
\end{equation}
 and the initial condition
 \begin{equation*}
  u(X;0) = \delta_Y(X) \hspace{0.2cm}\textrm{when}\hspace{0.2cm}x_1<\cdots<x_N \label{initial0}
 \end{equation*}
 is the transition probability from the initial configuration $Y$ to a configuration $X$ at time $t$. \\
  \indent Similarly, the forward equation of the drop-push model with $N$ particles is (\ref{totallyMaster}) and the boundary equation is
\begin{equation}
u(x_1,\cdots,x_i-1,x_i,\cdots,x_N;t) = u(x_1,\cdots,x_i,x_i,\cdots,x_N;t). \label{boundaryPUSH}
\end{equation}
 Alimohammadi et al. \cite{Ali,Ali2} combined (\ref{boundaryTASEP}) and (\ref{boundaryPUSH}) in the form of
\begin{eqnarray}
 & & u(x_1,\cdots,x_i,x_i,\cdots,x_N;t) \label{boundary}\\
  & &\hspace{0.5cm} = \mu u(x_1,\cdots,x_i-1,x_i,\cdots,x_N;t) + \lambda u(x_1,\cdots,x_i,x_i+1,\cdots,x_N;t)\nonumber
\end{eqnarray}
 and have shown that (\ref{boundary}) describes the process (\ref{pushingR}) with rates (\ref{RightRate}) or the process (\ref{BothPushing}) in the asymmetric case  with rates $pr_n$ and $ql_n$ given by (\ref{RightRate}) and (\ref{LeftRate1}).   It is possible to approach Alimohammadi et al.'s model in a different point of view. Let us  assume that  we have an interacting particle system  of which process is (\ref{BothPushing}) with rates $pr_n$ and $ql_n$ to the right and to the left, respectively. These rates are to be determined so that the model is  Bethe ansatz solvable. Let us consider the two-particle system.  Setting  $r_1=l_1=1$, the forward equation is  for $u(x_1,x_2;t)$ on $\mathbb{Z}^2 \times [0, \infty)$
 \begin{eqnarray}
  \label{master} \frac{d}{dt}~u(x_1,x_2;t) &=& pu(x_1-1,x_2;t) + pu(x_1,x_2-1;t)\\
                               & & + qu(x_1+1,x_2;t) + qu(x_1,x_2+1;t)-2u(x_1,x_2;t),\nonumber
 \end{eqnarray}
 and the boundary condition is
 \begin{eqnarray}
& &pr_1u(x,x;t) - p{r_2}u(x-1,x;t) - p(r_1-{r_2})u(x,x+1)\label{boundary2}\\
& =& -ql_1u(x+1,x+1;t) + q{l_2}u(x +1,x+2;t) + q(l_1-{l_2})u(x,x+1)\nonumber
\end{eqnarray}
for all $x \in \mathbb{Z}$. However, instead of (\ref{boundary2}), we consider a sufficient condition for it. To do so,  we set the right hand side and the left hand side of (\ref{boundary2}) equal to  zero, and letting  $\lambda = l_2/l_1, \mu =r_2/r_1$ and $\lambda+\mu =1$, then both equations reduce into one single equation
\begin{equation}
 u(x,x;t) = \mu u(x-1,x;t) + \lambda u(x,x+1;t). \label{boundary3}
\end{equation}
Hence, if $u(x_1,x_2;t)$ satisfies (\ref{boundary3}), it also satisfies (\ref{boundary2}).  This is the boundary condition suggested  in \cite{Ali,Ali2} as a combination of the boundary conditions of the TASEP and the drop-push model. Hence, we take (\ref{boundary}) as the boundary condition for a two-particle sector (when only two particles are adjacent) of an $N$-particle system. It has been shown \cite{Ali} that  (\ref{boundary}) implies the boundary conditions for a general $N$-particle system if $r_n$ and $l_n$ are given by (\ref{RightRate})  and (\ref{LeftRate1}), and so the model is Bethe ansatz solvable. Applying  (\ref{boundary3}) to the Bethe ansatz solution
 \begin{equation*}
 A_{12}\xi_1^{x_1}\xi_2^{x_2} + A_{21}\xi_1^{x_2}\xi_2^{x_1}
 \end{equation*}
 generates the $S$-matrix
\begin{equation}
S_{\beta\alpha} =-\frac{\xi_{\beta}}{\xi_{\alpha}}\cdot\frac{\mu + \lambda \xi_{\alpha}{\xi_{\beta}} - \xi_{\alpha}}{\mu + \lambda \xi_{\alpha}{\xi_{\beta}} - \xi_{\beta}}:=\frac{\xi_{\beta}}{\xi_{\alpha}} \cdot S_{\beta\alpha}^{\dag}. \label{SmatrixofPush}
\end{equation}
\\
\indent In the ASAP with two particles if $x_1 < x_2-1$, then the forward equation  is  (\ref{master}),    and if $x_1=x_2-1=x$, the forward equation is
\begin{eqnarray}
  \frac{d}{dt}~u(x,x+1;t) &=& pu(x-1,x+1;t) + qu(x,x +2;t) \label{master0} \\
                               & & + \lambda(p+q\mu)\sum_{n=1}^{\infty}\mu^{n-1}u(x-n,x-n+1;t) \nonumber\\
                               & & - u(x,x+1;t)-(p+q\mu)u(x,x+1;t). \nonumber
\end{eqnarray}
Thus the boundary condition
 \begin{eqnarray}
 & & pu(x,x;t) - p\lambda \sum_{n=1}^{\infty}\mu^{n-1}u(x-n,x-n+1;t)  \label{boundary6}\\
 & &\hspace{0.5cm} = \nonumber  - qu(x+1,x+1;t) + q\lambda \sum_{n=1}^{\infty}\mu^{n}u(x-n,x-n+1;t)
\end{eqnarray}
is obtained by subtracting (\ref{master0}) from (\ref{master}) with $x_1=x,x_2=x+1$. However, setting both sides of (\ref{boundary6}) to be zero, (\ref{boundary6}) is satisfied if $u(x_1,x_2;t)$ satisfies
 \begin{equation}
  u(x,x;t) = \lambda\sum_{n=0}^{\infty}\mu^n u(x-n-1,x-n;t).\label{boundary7}
 \end{equation}
It can be shown that (\ref{boundary7}) is equivalent to
 \begin{equation}
  u(x,x;t) = \lambda u(x-1,x;t) + \mu u(x-1,x-1;t) \label{boundary10}
 \end{equation}
 by applying (\ref{boundary10}) to itself recursively.
Hence we take
\begin{eqnarray}
 & &u(x_1,\cdots,x_i,x_i,\cdots,x_N;t) \label{boundary8}\\
 & & ~~= \lambda u(x_1,\cdots,x_i-1,x_i,\cdots,x_N;t) + \mu u(x_1,\cdots,x_i-1,x_i-1,\cdots,x_N;t) \nonumber
\end{eqnarray}
as the boundary condition for a two-particle sector of the ASAP with $N$ particles, and the resulting  $S$-matrix is
\begin{equation}
S_{\beta\alpha} =-\frac{\mu + \lambda\xi_{\beta} -\xi_{\beta}\xi_{\alpha}}{\mu + \lambda\xi_{\alpha} -\xi_{\beta}\xi_{\alpha}}. \label{SmatrixofASAP}
\end{equation}
The boundary condition (\ref{boundary8}) implies the boundary conditions for $N$-particle interactions when $\mu_n$ is given by (\ref{Constraint}) \cite{Povol1,Povol2}.\\
\indent
 In case of the AZRP with the assumptions mentioned in section 1, the $S$-matrix generated by the boundary condition
\begin{eqnarray}
 & &u(x_1,\cdots,x_i,x_i,\cdots,x_N) \label{boundary9}\\
 & & ~~= pu(x_1,\cdots,x_i,x_i-1,\cdots,x_N) + qu(x_1,\cdots,x_i+1,x_i,\cdots,x_N) \nonumber
\end{eqnarray}
is
\begin{equation*}
S_{\beta\alpha} =-\frac{\xi_{\alpha}}{\xi_{\beta}}\cdot\frac{p + q \xi_{\alpha}{\xi_{\beta}} - \xi_{\beta}}{p + q \xi_{\alpha}{\xi_{\beta}} - \xi_{\alpha}}:=\frac{\xi_{\alpha}}{\xi_{\beta}}\cdot S_{\beta\alpha}^{\ddag}. \label{SmatrixofAZRP}
\end{equation*}
 \indent Setting up the problem with $N$ particles, for all three models we have the forward equation
   \begin{eqnarray}
 \frac{d}{dt}~u(X;t) &=& \sum_{i=1}^N\Big(pu(x_1,\cdots,x_{i-1},x_i-1,x_{i+1},\cdots,x_N;t) \label{PushASEPMaster}\\
                                 & &\hspace{0.5cm}+~ qu(x_1,\cdots,x_{i-1},x_i+1,x_{i+1},\cdots,x_N;t) -~ Nu(X;t)\Big)  \nonumber
\end{eqnarray}
and the initial condition
\begin{equation}
 u(X;0) = \delta_Y(X) \label{initialCon}
\end{equation}
 which holds in the physical region of the corresponding model, and finally, the boundary conditions (\ref{boundary}), (\ref{boundary8}) and (\ref{boundary9}) for the two-sided PushASEP, the ASAP and the AZRP, respectively.

\section{Transition Probabilities for  $N$-particle system}
  In case of the two-sided PushASEP, the transition probability was expressed \cite{Ali2} as an integral over the unit circle, and thus $S$-matrices in the integrand are not exactly in the form of (\ref{SmatrixofPush}) to satisfy the initial condition. However, as a first step  to compute time-integrated currents we need the integral representation over a contour with sufficiently small radius or over a contour with sufficiently large radius so that a certain geometric series converges \cite{TW1}. So, in this section we find  the transition probabilities in the form of the one of the ASEP in \cite{TW1}. \\
\indent As usual,  the \textit{energy} in the Bethe ansatz solution is $\varepsilon_N(\xi) = \sum_i^N \varepsilon(\xi_i)$ where
\begin{equation*}
 \varepsilon (\xi_i) =  \frac{p}{\xi_i} + q \xi_i -1
\end{equation*}
and let
\begin{equation*}
 A_{\sigma}= \prod_{(\beta,\alpha)} S_{\beta\alpha} \label{product1}
\end{equation*}
where the product is over all inversions $(\beta, \alpha)$ in a permutation $\sigma \in \mathbb{S}_N$ and $S_{\beta\alpha}$ is the $S$-matrix of the corresponding model. Also, we assume that a differential $d\xi$ incorporates $\frac{1}{2\pi i}$.
\begin{theorem}
 The transition probabilities of the two-sided PushASEP, the ASAP and the AZRP are given in the form of
\begin{equation}
 P_Y(X;t) = \sum_{\sigma \in \mathbb{S}_N} \int_{\mathcal{C}_r}\cdots \int_{\mathcal{C}_r} A_{\sigma} \prod_i \xi_{\sigma(i)}^{x_i - y_{\sigma(i) }-1} e^{\sum_i \varepsilon (\xi_i)t} d\xi_1 \cdots d\xi_N, \label{transition}
 \end{equation}
 where $\mathcal{C}_r$ is a circle centered at zero with sufficiently small radius $r$ so that all the poles  of the integrand except at the origin lie outside $\mathcal{C}_r$.
\end{theorem}
 \noindent  The proof of the theorem consists of showing that (\ref{transition})  satisfies ($\rmnum{1}$) the forward equation (\ref{PushASEPMaster}) for all $X \in \mathbb{Z}^N$, ($\rmnum{2}$) the corresponding boundary conditions (\ref{boundary}),(\ref{boundary8}) or (\ref{boundary9}), and ($\rmnum{3}$) the initial condition (\ref{initialCon}) in the physical region of the corresponding model. The proofs  of ($\rmnum{1}$) and  ($\rmnum{2}$)  are the same as the proof of the ASEP case \cite{TW1}.   For
 ($\rmnum{3}$) it suffices to show that
 \begin{equation*}
  \sum_{\sigma \neq id} \int_{\mathcal{C}_r}\cdots \int_{\mathcal{C}_r} A_{\sigma} \prod_i \xi_{\sigma(i)}^{x_i - y_{\sigma(i) }-1} d\xi_1 \cdots d\xi_N =0. \label{formula1}
 \end{equation*}
 In case of the ASAP, the $S$-matrix of the ASAP (\ref{SmatrixofASAP}) is in the  same form  as the $S$-matrix of the ASEP up to constants, hence the proof is the same as the proof of the ASEP case \cite{TW1}.  The $S$-matrix of the AZRP has an additional factor besides the $S$-matrix of the ASEP and the $S$-matrix of the two-sided PushASEP is in the reciprocal form of the $S$-matrix of the AZRP. Hence it is natural to use the techniques of the proofs of relevant lemmas in \cite{TW1} in order to prove ($\rmnum{3}$). Let
 \begin{equation*}
  I(\sigma):= \int_{\mathcal{C}_r}\cdots \int_{\mathcal{C}_r}A_{\sigma} \prod_i \xi_{\sigma(i)}^{x_i - y_{\sigma(i) }-1} d\xi_1 \cdots d\xi_N.
 \end{equation*}
Now we present some lemmas corresponding to Lemma 2.1 -- Lemma 2.4 in \cite{TW1}.
\begin{lemma}
  Suppose that $\beta$ appears in position $\beta +1$ in $\sigma$ and the entries following  $\beta$ are greater than $\beta$.  Then $I(\sigma) = 0$ for the two-sided PushASEP.
\end{lemma}
\begin{proof}
With the assumption the permutation $\sigma$ has a unique inversion of the form $(\alpha,\beta)$. Let $\gamma$ be the position of $\alpha$. There are  possibly other inversions of the form $(\alpha,\delta)$ but no other types of inversions.  The number of inversions of the form $(\alpha ,\delta)$ is $\alpha -\gamma$ including $(\alpha,\beta)$. In this case, the variables $\xi_{\alpha}$ and $\xi_{\beta}$ in the integrand appear as
 \begin{equation*}
  \prod_{(\alpha,\delta)} S_{\alpha\delta}^{\dag}\xi_{\alpha}^{x_{\gamma}-y_{\alpha}  -1 + \alpha - \gamma}\xi_{\beta}^{x_{\beta+1} - y_{\beta} -2}
\end{equation*}
where the product is over all inversions of the form $(\alpha ,\delta)$.
Substituting
\begin{equation}
\xi_{\beta} = \frac{\eta}{\prod_{\delta \neq \beta}\xi_{\delta}},\label{sub1}
\end{equation}
in the integrand and in the differential, the variable $\xi_{\alpha}$ appears as
\begin{equation*}
 \xi_{\alpha}^{x_{\gamma} - x_{\beta+1} + y_{\beta} - y_{\alpha} +\alpha -\gamma}
\end{equation*}
and the exponent of $\xi_{\alpha}$ satisfies
\begin{equation}
x_{\gamma} - x_{\beta+1} + y_{\beta} - y_{\alpha} +\alpha -\gamma \leq -1 \label{exponent1}
\end{equation}
 because $  x_{\beta+1} - x_{\gamma}  \geq (\beta +1) -\gamma$ and $ y_{\alpha} - y_{\beta} \geq \alpha - \beta$. Also we have
\begin{eqnarray*}
 \prod_{(\alpha,\delta)} S_{\alpha\delta}^{\dag} & = &
(-1)^{\alpha-\gamma}\frac{\mu + \lambda \eta \prod_{\delta \neq \alpha,\beta}\xi_{\delta}^{-1} - \eta\prod_{\delta \neq \beta}\xi_{\delta}^{-1}}{\mu + \lambda \eta \prod_{\delta \neq \alpha,\beta}\xi_{\delta}^{-1} - \xi_{\alpha}} \\
& & \hspace{0.5cm} \times \prod_{\substack{(\alpha,\delta) \\ \delta \neq \beta }} \frac{\mu + \lambda \xi_{\alpha}\xi_{\delta} - \xi_{\delta}}{\mu + \lambda \xi_{\alpha}\xi_{\delta} - \xi_{\alpha}}
\end{eqnarray*}
by the substitution (\ref{sub1}). If we integrate with respect to $\xi_{\alpha}$ over $\mathcal{C}_r$, the order of the pole in $\eta\prod_{\delta \neq \beta}\xi_{\delta}^{-1}$ is greater than  2 because of (\ref{exponent1}).  Moreover, the singularities in the denominators are bounded away from zero because the second summands of denominators are $O(r^2)$ and $\mu \neq 0$, and thus   $I(\sigma) = 0$.
\end{proof}
\noindent The next lemma shows that Lemma 2.1 in \cite{TW1} holds for the AZRP in the physical region of the AZRP.
\begin{lemma} (The AZRP version of Lemma 2.1 in \cite{TW1})
  Suppose that $\beta$ appears in position $\beta -1$ in $\sigma$ and the entries preceding $\beta$ are less than $\beta$.  Then $I(\sigma) = 0$ for the AZRP.
\end{lemma}
\begin{proof}
With the assumption the permutation $\sigma$ has a unique inversion of the form $(\beta,\alpha)$.  Let $\gamma$ be the position of $\alpha$. There are  possibly other inversions of the form $(\delta,\alpha)$ but no other types of inversions. The number of inversions of the form $(\delta,\alpha)$ is $\gamma-\alpha$ including $(\beta,\alpha)$. In this case, the variables $\xi_{\alpha}$ and $\xi_{\beta}$ in the integrand appear as
 \begin{equation*}
  \prod_{(\delta,\alpha)} S_{\delta\alpha}^{\ddag}\xi_{\alpha}^{x_{\gamma}-y_{\alpha}  -1 +  \gamma-\alpha}\xi_{\beta}^{x_{\beta-1} - y_{\beta} -2}
\end{equation*}
where the product is over all inversions of the form $(\delta,\alpha)$.
Substituting
\begin{equation*}
\xi_{\beta} = \frac{\eta}{\prod_{\delta \neq \beta}\xi_{\delta}},
\end{equation*}
in the integrand and in the differential, the variable $\xi_{\alpha}$ appears as
\begin{equation*}
 \xi_{\alpha}^{x_{\gamma} - x_{\beta-1} + y_{\beta} - y_{\alpha} +\gamma-\alpha}
\end{equation*}
and the exponent of $\xi_{\alpha}$ satisfies
\begin{equation*}
x_{\gamma} - x_{\beta-1} + y_{\beta} - y_{\alpha} +\gamma -\alpha \geq 1 \label{exponent2}
\end{equation*}
 because $  x_{\gamma} - x_{\beta-1}  \geq 0$,  $  y_{\beta} - y_{\alpha} \geq 0$ and $\gamma - \alpha  \geq 1$. The rest of the proof is the same as the proof of Lemma 2.1. in \cite{TW1}.
\end{proof}
\begin{lemma} Let $\sigma \in \mathbb{S}_N$ and $(\beta,\alpha)$ be an inversion of $\sigma \in \mathbb{S}_N$. Then,
\begin{equation*}
  \prod_{(\beta,\alpha)} \frac{\xi_{\beta}}{\xi_{\alpha}} = \prod_i\xi_{\sigma(i)}^{\sigma(i)-i}.
\end{equation*}
\end{lemma}
\begin{proof}  There are $N-i$ entries to the right  of $\sigma(i)$ and $i-1$ entries to the left of $\sigma(i)$ and $\sigma(i)-1$ entries are smaller than $\sigma(i)$. Suppose that there are $n$ inversions of the form $(\sigma(i),\sigma(j))$. This implies that the number of entries smaller than $\sigma(i)$ is $n$ and the $n$ entries lie to the right of $\sigma(i)$. Hence $\sigma(i)-1-n$ entries less than $\sigma(i)$ lie   to the left of $\sigma(i)$, which implies that $(i-1) - (\sigma(i)-1-n)$ entries are greater than $\sigma(i)$ and  lie to the left of  $\sigma(i)$.
\end{proof}
\begin{lemma}
 Let $x_i,y_i \in \mathbb{Z}$  and $(b,a)$ be an inversion of a permutation $\sigma \in \mathbb{S}_N$. Define
 \begin{equation*}
  h_{\sigma}(\xi_1,\cdots,\xi_N):= \prod_{(b,a)} \frac{\xi_{b}}{\xi_{a}} \prod_i\xi_{\sigma(i)}^{x_i - y_{\sigma(i)} -1}
 \end{equation*}
  and denote by $h_{\sigma}^{\gamma}$ the function obtained by substituting $\xi_{\gamma} = \frac{\eta}{\prod_{\mu \neq \gamma} \xi_{\mu}}$ in the function $h_{\sigma}$. Assume that $x_i <x_{i+1}$ and $y_i <y_{i+1}$ for all $i$. ($\rmnum{1}$)
 If two permutations $\sigma$ and $\sigma'$ differ only by an interchange of two adjacent entries $\alpha$ and $\beta$, then $h_{\sigma}^{\gamma}|_{\xi_{\alpha} = \xi_{\beta}} = h_{\sigma'}^{\gamma}|_{\xi_{\alpha} = \xi_{\beta}}$ when $\gamma \neq \alpha, \beta$. $(\rmnum{2})$  Moreover, if $\gamma$ is to the right of $\alpha$ and $\beta$  and $\gamma < \alpha,\beta$, then the largest power of $\xi_{\alpha}$ and $\xi_{\beta}$ in $h_{\sigma}^{\gamma}$ and $h_{\sigma'}^{\gamma}$ is zero.
\end{lemma}
\begin{proof}
Let $\alpha < \beta$. Assuming that  $\sigma(i) = \alpha$, $\sigma(i+1) =\beta$,  $\sigma'(i+1) = \alpha$, $\sigma'(i) =\beta$ and $\sigma(j)=\sigma'(j)=\gamma$. Then by Lemma 3.4 the variables $\xi_{\alpha}$,  $\xi_{\beta}$ and $\xi_{\gamma}$   appear as
\begin{equation*}
  \xi_{\alpha}^{x_i - y_{\alpha} -1 + \alpha -i}\xi_{\beta}^{x_{i+1} - y_{\beta} -1 + \beta - (i+1)}\xi_{\gamma}^{x_j -y_{\gamma}-1 + \gamma -j} \label{variables}
\end{equation*}
in $h_{\sigma}$  and   appear  as
\begin{equation*}
  \xi_{\alpha}^{x_{i+1} - y_{\alpha} -1 + \alpha -(i+1)}\xi_{\beta}^{x_{i} - y_{\beta} -1 + \beta - i}\xi_{\gamma}^{x_j -y_{\gamma}-1 + \gamma -j}\label{variables2}
\end{equation*}
in $h_{\sigma'}$.
By the substitution ($\gamma \neq \alpha,  \beta$), the variables $\xi_{\alpha}$ and $\xi_{\beta}$ appear as
\begin{equation*}
   \xi_{\alpha}^{x_i - x_j + y_{\gamma} - y_{\alpha} +\alpha - \gamma +j-i}\xi_{\beta}^{x_{i+1} - x_j + y_{\gamma} - y_{\beta} + \beta-\gamma + j-(i+1) } \label{variables33}
\end{equation*}
in $h_{\sigma}^{\gamma}$ and
\begin{equation*}
  \xi_{\alpha}^{x_{i+1} - x_j + y_{\gamma} - y_{\alpha} +\alpha -\gamma + j-(i+1) }\xi_{\beta}^{x_{i} - x_j + y_{\gamma} - y_{\beta} +\beta- \gamma +j-i}\label{variables4}
\end{equation*}
in $h_{\sigma'}^{\gamma}$.  Thus,    ($\rmnum{1}$) follows from the fact that the sum of exponents of  $\xi_{\alpha}$ and $\xi_{\beta}$ in $h_{\sigma}^{\gamma}$ is equal to the sum of exponents of  $\xi_{\alpha}$ and $\xi_{\beta}$ in $h_{\sigma'}^{\gamma}$. Now, suppose that $j>i+1$ and $\gamma < \alpha, \beta$. Then  $(\rmnum{2})$ follows from the fact that $x_i < x_{i+1} < x_j$ and $y_{\gamma} < y_{\alpha} < y_{\beta}$.
\end{proof}
\noindent The next lemma is for the AZRP. The proof is almost the same as the proof of Lemma 3.5 and so we omit the proof.
\begin{lemma}
 Let $x_i,y_i \in \mathbb{Z}$  and $(b,a)$ be an inversion of a permutation $\sigma \in \mathbb{S}_N$. Define
 \begin{equation*}
  h_{\sigma}(\xi_1,\cdots,\xi_N):= \prod_{(b,a)} \frac{\xi_{a}}{\xi_{b}} \prod_i\xi_{\sigma(i)}^{x_i - y_{\sigma(i)} -1}
 \end{equation*}
  and denote by $h_{\sigma}^{\gamma}$ the function obtained by substituting $\xi_{\gamma} = \frac{\eta}{\prod_{\mu \neq \gamma} \xi_{\mu}}$ in the function $h_{\sigma}$.  Assume that $x_i \leq x_{i+1}$ and $y_i \leq y_{i+1}$ for all $i$. $(\rmnum{1})$
 If two permutations $\sigma$ and $\sigma'$ differ only by an interchange of two adjacent entries $\alpha$ and $\beta$, then $h_{\sigma}^{\gamma}|_{\xi_{\alpha} = \xi_{\beta}} = h_{\sigma'}^{\gamma}|_{\xi_{\alpha} = \xi_{\beta}}$ when $\gamma \neq \alpha, \beta$. $(\rmnum{2})$ Moreover, if $\gamma$ is to the left $\alpha$ and $\beta$ and $\gamma > \alpha,\beta$, then the powers of $\xi_{\alpha}$ and $\xi_{\beta}$ in $h_{\sigma}^{\gamma}$ and $h_{\sigma'}^{\gamma}$ are greater than or equal to 2.
\end{lemma}
\begin{lemma} Suppose that in the permutations $\sigma$ and $\sigma'$ the entry $\gamma$ appears to the right of two adjacent entries $\alpha,\beta> \gamma$, and that the permutations differ only by an interchange of $\alpha$ and $\beta$. Then $I(\sigma) + I(\sigma') = 0$ for the two-sided PushASEP.
\end{lemma}
\begin{proof}
We will show that $I(\sigma) + I(\sigma') = 0$ when we integrate with respect to $\xi_{\alpha}$ and $\xi_{\beta}$.
 Suppose that $\alpha < \beta$, $\alpha = \sigma(i), \beta = \sigma(i+1)$ and $\gamma = \sigma(j)$ with $j>i+1$. From the assumption there are inversions $(\alpha,\gamma)$ and $(\beta,\gamma)$ in both $\sigma$ and $\sigma'$ and  there is additionally $(\beta,\alpha)$ in $\sigma'$. For $\sigma$ the variables $\xi_{\alpha}, \xi_{\beta}$ and $\xi_{\gamma}$ appear as
 \begin{equation}
 \xi_{\alpha}^{\alpha -i}\xi_{\beta}^{\beta -(i+1)}\xi_{\gamma}^{\gamma -j}S_{\alpha\gamma}^{\dag}S_{\beta\gamma}^{\dag}\prod S_{\delta\delta'}^{\dag}\xi_{\alpha}^{x_i -y_{\alpha} -1}\xi_{\beta}^{x_{i+1} -y_{\beta} -1}\xi_{\gamma}^{x_j -y_{\gamma} -1} \label{eq1}
 \end{equation}
 in the integrand where the product is over all the other inversions involving ${\alpha},{\beta}$ and ${\gamma}$ excluding $(\alpha,\gamma)$ and $(\beta,\gamma)$. All the $S$-matrices in $\sigma$ appear in  $\sigma'$, too. Substituting  $\xi_{\gamma} = \frac{\eta}{\prod_{\delta \neq \gamma}\xi_{\delta}}$ in the integrand and in the differential,  the powers of  $\xi_{\alpha}$ and $\xi_{\beta}$ are less than or equal to $-1$ from Lemma 3.5. The product $S_{\alpha\gamma}^{\dag}S_{\beta\gamma}^{\dag}$ is
 \begin{equation*}
 \frac{\mu + \lambda \eta\prod_{\delta \neq \alpha,\gamma}\xi_{\delta}^{-1} - \eta\prod_{\delta \neq \gamma} \xi_{\delta}^{-1}}{\mu + \lambda \eta\prod_{\delta \neq \alpha,\gamma}\xi_{\delta}^{-1} - \xi_{\alpha}}
 \frac{\mu + \lambda \eta\prod_{\delta \neq \beta,\gamma}\xi_{\delta}^{-1} - \eta\prod_{\delta \neq \gamma} \xi_{\delta}^{-1}}{\mu + \lambda \eta\prod_{\delta \neq \beta,\gamma}\xi_{\delta}^{-1} - \xi_{\beta}}, \label{product}
 \end{equation*}
  which is in the same form as the one in Lemma 2.2 in \cite{TW1}. Thus, with the same argument as in Lemma 2.2 in \cite{TW1} we have $\xi_{\alpha} = \xi_{\beta}$. Similarly, for $\sigma'$ the variables $\xi_{\alpha}, \xi_{\beta}$ and $\xi_{\gamma}$ appear as
 \begin{equation}
 \xi_{\alpha}^{\alpha -(i+1)}\xi_{\beta}^{\beta -i}\xi_{\gamma}^{\gamma -j}S_{\beta\alpha}^{\dag}S_{\alpha\gamma}^{\dag}S_{\beta\gamma}^{\dag}\prod S_{\delta\delta'}^{\dag}\xi_{\beta}^{x_i -y_{\beta} -1}\xi_{\alpha}^{x_{i+1} -y_{\alpha} -1}\xi_{\gamma}^{x_j -y_{\gamma} -1} \label{eq2}
 \end{equation}
 in the integrand. Substituting $\xi_{\gamma} = \frac{\eta}{\prod_{\delta \neq \gamma}\xi_{\delta}}$, it follows from Lemma 3.5 that  (\ref{eq1}) is the negative of (\ref{eq2}) when  $\xi_{\alpha} = \xi_{\beta}$, because $S_{\beta\alpha}^{\dag}=-1$ when $\xi_{\alpha} = \xi_{\beta}$.
Hence  the integrands in $I(\sigma)$ and $I(\sigma')$ differ only by the sign after integrating with respect to $\xi_{\alpha}$ and $\xi_{\beta}$ and so $I(\sigma) + I(\sigma') = 0$.
 \end{proof}
 \noindent The following lemma can be proved by using Lemma 3.6 and the same procedure as in the proof of Lemma 3.7, and hence we omit the proof.
 \begin{lemma} (AZRP version of Lemma 2.2 in \cite{TW1}) Suppose that in the permutations $\sigma$ and $\sigma'$ the entry $\gamma$ appears to the left of two adjacent entries $\alpha,\beta < \gamma$, and that the permutations differ only by an interchange of $\alpha$ and $\beta$. Then $I(\sigma) + I(\sigma') = 0$ for the AZRP.
\end{lemma}

\noindent While Lemma 2.4 in \cite{TW1} is for the ASEP and the AZRP, the next lemma is for the two-sided PushASEP.
\begin{lemma}(The two-sided PushASEP version of Lemma 2.4 in \cite{TW1}) For any $N$ the set $\mathbb{S}_N \setminus \{\textrm{id}\}$ is the union of disjoint subsets, each of which consists of either of a single permutation satisfying Lemma 3.2 or a pair of permutations satisfying Lemma 3.7.
\end{lemma}
\begin{proof} It is clear that the result holds for $N=2$. Assume that the result holds for $N-1$. Let $A_{\alpha}$ be the set of permutations in which $\alpha$ appears in position $N$. If $\alpha = N$, we apply the induction. Notice that all permutations in $A_{N-1}$ satisfy Lemma 3.2. If $\alpha < N-1$, we can pair $(N-1)!$ permutations in $A_{\alpha}$ as in Lemma 2.3 in \cite{TW1}, and these pairs satisfy Lemma 3.7.
\end{proof}
\noindent Hence, for the two-sided PushASEP the proof of ($\rmnum{3}$) of Theorem 3.1  follows from Lemma 3.2, 3.7, and 3.9  and for the AZRP the proof follows  from Lemma 3.3, 3.8 and Lemma 2.4 in \cite{TW1}. This completes the proof of Theorem 3.1.
\\
\indent An immediate result of Theorem 3.1 is a natural mapping between the state spaces of the ASEP and the AZRP under which transition probabilities are conserved. This can be considered as another approach of the well-known mapping between  the ASEP and the AZRP \cite{Evans,Harris,Kipnis2,Timo}. Specifically, let $S$ and $S'$ be countable state spaces  of the ASEP and the AZRP with $N$ particles, respectively, and $X,Y \in S$ and $X',Y' \in S'$.  Let $A_{\sigma}^A$ and $A_{\sigma}^{Z}$ be coefficients of Bethe ansatz solutions of the ASEP and the AZRP, respectively. Then under the bijection $f$ from $S'$ to $S$ defined by
\begin{equation}
 f:  (x_i)_i\rightarrow (x_i +i)_i, \label{bijection}
\end{equation}
we have from Lemma 3.4
\begin{eqnarray*}
 P_Y'(X';t) &=& \sum_{\sigma \in \mathbb{S}_N} \int_{\mathcal{C}_r}\cdots \int_{\mathcal{C}_r} A_{\sigma}^Z \prod_i \xi_{\sigma(i)}^{x'_i- y'_{\sigma(i) } -1} e^{\sum_i \varepsilon (\xi_i)t} d\xi_1 \cdots d\xi_N \\
          &=& \sum_{\sigma \in \mathbb{S}_N} \int_{\mathcal{C}_r}\cdots \int_{\mathcal{C}_r} A_{\sigma}^{A} \prod_i \xi_{\sigma(i)}^{x'_i +i- y'_{\sigma(i) } - \sigma(i)-1} e^{\sum_i \varepsilon (\xi_i)t} d\xi_1 \cdots d\xi_N \\
          &=&\sum_{\sigma \in \mathbb{S}_N} \int_{\mathcal{C}_r}\cdots \int_{\mathcal{C}_r} A_{\sigma}^A \prod_i \xi_{\sigma(i)}^{x_i- y_{\sigma(i) } -1} e^{\sum_i \varepsilon (\xi_i)t} d\xi_1 \cdots d\xi_N\\
          &=& P_{Y}(X;t).
\end{eqnarray*}
Hence  the distributions of the $m$th particle's position from the leftmost in the ASEP and in the AZRP  are conserved under the mapping  (\ref{bijection}) because  the distributions of the $m$th particle's position are obtained by summing transition probabilities over all possible configurations. Thus, $\mathbb{P}(x_m(t) =x)$, the probability of finding the $m$th particle at $x$ at time $t$ in the AZRP  is given by (5.12) in \cite{TW1} with $y_i$ replaced by $y_i +i$. Let us modify the $N$-dimensional integrand (1.1) in \cite{TW1} by letting
\begin{equation*}
 I_{Z}(x,Y,\xi) = \prod_{i<j}\frac{\xi_j - \xi_i}{p + q \xi_i\xi_j -\xi_i}\frac{1-\xi_1\cdots \xi_N}{(1-\xi_1)\cdots(1-\xi_N)}\prod_i\Big(\xi_i^{x-(y_i +i)-1}e^{\varepsilon(\xi_i)t}\Big)
\end{equation*}
and  as in \cite{TW1}, denote by $I_Z(x,Y_S,\xi)$ the integrand analogous to $I_Z(x,Y,\xi)$, where only the variables $\xi_i$ with $i \in S$ occur. Then we can immediately write the explicit form of
$\mathbb{P}(x_m(t) =x)$ for the AZRP.
\begin{proposition} Let $S \subset \{1,\cdots,N\}$ and $\sigma(S) = \sum_{i \in S}i$. Then
\begin{eqnarray}
   \mathbb{P}(x_m(t) =x) &=& (-1)^{m+1}(pq)^{m(m-1)/2} \label{mth} \\
                         & & \times \sum_{|S|\geq m} {|S| -1 \brack |S| -m}\frac{p^{\sigma(S) -m|S|}}{q^{\sigma(S)-|S|(|S|+1)/2}}\int_{\mathcal{C}_R}\cdots \int_{\mathcal{C}_R} I_Z(x,Y_S,\xi)d^{|S|}\xi \nonumber
\end{eqnarray}
where $\mathcal{C}_R$ is a circle centered at 0 with sufficiently large radius $R$ so that the poles of the integrand are in the circle.
\end{proposition}
\noindent It is natural to expect that $\mathbb{P}(x_m(t) = x)$ in the ASEP with step initial condition should be equal to that in the AZRP with the \textit{delta function-like} initial condition that infinitely many particles occupy the origin and all other sites are empty. This can be simply verified as follows. Let $S = \{z_1,\cdots, z_k\}$ in (\ref{mth}). As in the proof of Corollary in \cite{TW1}, by the change of variables, the factor
$$\prod_i\Big(\xi_{z_i}^{x-(y_{z_i} +z_i)-1}e^{\varepsilon(\xi_{z_i})t}\Big)$$
in the integrand  $I_Z(x,Y_S,\xi)$ becomes
$$\prod_i\Big(\xi_i^{x-z_i-1}e^{\varepsilon(\xi_{i})t}\Big)$$
for the \textit{delta function-like} initial condition. Hence the rest of the proof is the same as in the proof of Corollary in \cite{TW1}. Moreover, the AZRP with the \textit{delta function-like} initial condition has the same asymptotic behavior as the ASEP with step initial condition as well as the Fredholm determinant representation \cite{TW2,TW3}.\\ \\
\noindent\textit{Remark 1.} The $S$-matrices of the ASAP and the ASEP are equivalent up to constants. Specifically, if $p$ and $q$ in the $S$-matrix of the ASEP are replaced by $-\mu/\lambda $ and $1/\lambda$, respectively, we obtain the $S$-matrix of the ASAP. Hence the integral formulas for the ASAP are obtained from the integral formulas of the ASEP with $p=-\mu/\lambda $ and $q=1/\lambda$ and without any change in the \textit{energy} term,  $\varepsilon(\xi_i)$.

\noindent\textit{Remark 2.} While there is a direct mapping between the ASEP and the AZRP, on the other hand, there is no such mapping between the ASEP and the two-sided PushASEP. Moreover, there is no such simple relation between the $S$-matrix of the PushASEP and the $S$-matrix of the ASEP as the ralation between the $S$-matrix of the ASAP and the $S$-matrix of the ASEP. Thus, it needs further investigation to study the current distribution and its asymptotics in the two-sided PushASEP and they remain as problems for the future.

\noindent\textbf{Acknowledgement}\\
The author is grateful to Craig A. Tracy for invaluable comments and supports on this work and also would like to thank Timo Sepp\"{a}l\"{a}inen for commenting on the ZRP by email and during the workshop at MSRI. This work was supported by National Science Foundation through the grant DMS-0906387.
\\

\end{document}